\documentclass[
 reprint,
 amsmath,amssymb,
 aps,
]{revtex4-1}
\usepackage{bm}
\usepackage[utf8]{inputenc}
\usepackage[T1]{fontenc}
\usepackage{mathptmx}

\usepackage[dvipsnames]{xcolor}
\usepackage{soul}
\usepackage{stmaryrd}
\usepackage[normalem]{ulem}
\usepackage{graphicx}
\usepackage[nice]{nicefrac}

\DeclareMathOperator\erf{erf}

\draft 

\newcommand{\ie}{\textit{i}.\textit{e}., }

\begin{document}

\title{An efficient Kinetic Monte Carlo to study analyte capture by a nanopore: \\ Transients, boundary conditions and time-dependent fields \\
} 

\author{Le Qiao}
\author{Maxime Ignacio}
\author{Gary W. Slater}
 \email{gslater@uottawa.ca}

\affiliation{Department of Physics, University of Ottawa, Ottawa, Ontario K1N 6N5, Canada}

\date{\today}

\begin{abstract}
To better understand the capture process by a nanopore, we introduce an efficient Kinetic Monte Carlo (KMC) algorithm that can simulate long times and large system sizes by mapping the dynamic of a point-like particle in a 3D spherically symmetric system onto the 1D biased random walk. Our algorithm recovers the steady-state analytical solution and allows us to study time-dependent processes such as transients. Simulation results show that the steady-state depletion zone near pore is barely larger than the pore radius and narrows at higher field intensities; as a result, the time to reach steady-state is much smaller than the time required to empty a zone of the size of the capture radius $\lambda_e$. When the sample reservoir has a finite size, a second depletion region propagates inward from the outer wall, and the capture rate starts decreasing when it reaches the capture radius $\lambda_e$. We also note that the flatness of the electric field near the pore, which is often neglected, induces a traffic jam that can increase the transient time by several orders of magnitude. Finally, we propose a new proof-of-concept scheme to separate two analytes of the same mobility but different diffusion coefficients using time-varying fields.
\end{abstract}


\maketitle 
\section{Introduction}
\label{section:intro}
Nanopore sensing based on voltage-driven translocation is a hot topic due to its potential applications for the analysis and detection of biomolecules\cite{beamishProgrammableDNANanoswitch2019,najafisohiDNACaptureNanopore2020,charronPreciseDNAConcentration2019,haganChemicallyTailoringNanopores2020a,bandarkarHowNanoporeTranslocation2020,heFastCaptureMultiplexed2019,shiNanoporeSensing2017,waughSolidstateNanoporeFabrication2020,garoliPlasmonicNanoporesSingleMolecule2019}. In short, an electric field is applied across a nanopore to electrophoretically capture and translocate charged analytes like DNA, RNA and proteins. Useful information about the analytes (\ie structure, type, length,  \textit{etc.})
can be obtained by analyzing ionic current modulations during the threading process. With more sophisticated nanopore fabrication technologies\cite{kimRapidFabricationUniformly2006,gilboaAutomatedUltraFast2020,garoliPlasmonicNanoporesSingleMolecule2019,waughSolidstateNanoporeFabrication2020,d.y.bandaraNanoporeSizingImproving2020} and growing understanding of the mechanisms controlling translocation\cite{sakaueNonequilibriumDynamicsPolymer2007,muthukumarPolymerTranslocation2011,wanunuNanoporesJourneyDNA2012,panjaEyeNeedleRecent2013,palyulinPolymerTranslocationFirst2014,buyukdagliTheoreticalModelingPolymer2019}, numerous novel systems have been designed to enhance performance and facilitate new applications\cite{haganChemicallyTailoringNanopores2020a,beamishProgrammableDNANanoswitch2019,briggsDNATranslocationsNanopores2018,lamEntropicTrappingDNA2019,charronPreciseDNAConcentration2019,najafisohiDNACaptureNanopore2020,eggenbergerFluidSurfaceCoatings2019,bhattacharyaTugWarDoublenanopore2020}. 
However, several aspects of the capture process are still poorly understood and difficult to observe directly, such as the depletion zone\cite{chenProbingSingleDNA2004,grosbergDNACaptureNanopore2010}, the capture radius\cite{qiaoVoltagedrivenTranslocationDefining2019,rowghanianElectrophoresisDNACoil2013,rowghanianElectrophoreticCaptureDNA2013,grosbergDNACaptureNanopore2010,nomidisDNACaptureClyA2018,nakaneEvaluationNanoporesCandidates2002,muthukumarTheoryCaptureRate2010}, non-equilibrium dynamics\cite{muthukumarTheoryCaptureRate2010,farahpourChainDeformationTranslocation2013,vollmerTranslocationNonequilibriumProcess2016,qiaoCaptureRodlikeMolecules2020,katkarRoleNonequilibriumConformations2018} and molecule-pore interactions\cite{buyukdagliControllingPolymerCapture2017}. High capture rates and large capture zones remain the most important criteria for an ideal (DC-field) translocation setup. Therefore, clarifying those aspects is essential to provide guidance for more precise control and better design of capture devices. Indeed, these goals have motivated the development of theoretical approaches and simulation models to study the interplay between diffusion and electrostatic forces during capture\cite{qiaoVoltagedrivenTranslocationDefining2019,nakaneEvaluationNanoporesCandidates2002,rowghanianElectrophoreticCaptureDNA2013,rowghanianElectrophoresisDNACoil2013,grosbergDNACaptureNanopore2010,muthukumarTheoryCaptureRate2010,nomidisDNACaptureClyA2018,muthukumarCommunicationChargeDiffusion2014,nakaneEvaluationNanoporesCandidates2002,katkarRoleNonequilibriumConformations2018,farahpourChainDeformationTranslocation2013,vollmerTranslocationNonequilibriumProcess2016,buyukdagliControllingPolymerCapture2017,qiaoCaptureRodlikeMolecules2020}. However, most of these theoretical studies focus solely on steady-state quantities while using the point-charge approximation for the field near the pore. We previously studied the time dependence of the capture process with Lattice Monte Carlo simulations and discussed the nature of the capture radius for point-like particles \cite{qiaoVoltagedrivenTranslocationDefining2019}. We also examined the impact of field-driven orientation on the capture of rod-like molecules using Langevin Dynamic (LD) simulations\cite{qiaoCaptureRodlikeMolecules2020}. Unfortunately, these previous algorithms and approaches do not allow us to easily study large 3D systems over long periods of time in the presence of time-varying fields and complex boundary conditions.

Our main goal is thus to introduce a new and flexible Kinetic Monte Carlo (KMC) algorithm that can efficiently study large drift-diffusion problems under a wide variety of conditions. We do this by mapping (or projecting) the dynamics of a point-like particle in d-dimensions onto a 1D biased random walk. This projection allows us to use a smaller lattice sizes and/or larger systems, in order to investigate short-time transients, the approach to steady state, the impact of time-dependent fields, and the long-time effect of finite size boundary conditions. As an example, we use this new tool to study the impact of realistic field lines near the pore as well as a novel idea to modify the relative concentration of co-migrating analytes in a mixture.

This paper is organized as follows: We first review the basic theoretical considerations of the capture problem and introduce the relevant time/length scales. Section~\ref{section: KMC} then describes the mapping from 3D to 1D including the corresponding KMC algorithm and the related boundary conditions. In the result section (\ref{section:results}), we first test our algorithm with the time-dependent concentration profiles and capture rates for the classical case of diffusion-limited absorption by a sphere. We then add the electric field in order to study nanopore capture in both finite and infinite systems. We also examine the reverse process of escape under opposite polarity conditions. Finally, we propose and briefly test a new pulsed-field concept that could separate a binary mixture in Sec.~\ref{section:separation}. We conclude the paper in Sec.~\ref{sec:discussionandcon} with a discussion of our main results, especially the impact of the flat field near the pore, and the potential applications of time-varying electric fields.

\section{Basic theoretical elements}
\label{section:theory}

As discussed in the Introduction, we assume spherical symmetry in the following. The drift-Diffusion equation for the (point-like) analyte concentration $C(r,t)$ in three dimensions and in spherical coordinates is given by\cite{nomidisDNACaptureClyA2018,grosbergDNACaptureNanopore2010}

\begin{equation}
\label{eq:diff}
\frac{\partial C(r,t)}{\partial t} \!  = \! \frac{D}{r^{2}} \frac{\partial }{\partial r} \! \left[ r^{2} \left( \frac{\partial C(r,t)}{\partial r} ~ + ~ \frac{C(r,t)}{k_BT} 
\frac{\partial U_e}{\partial r}\right)\right],
\end{equation}
where $D$ is its diffusion coefficient and $U_e(r)$ is the electrostatic potential energy at radial position $r$.

Theoretical studies of analyte capture generally use the point-charge field (PCF) approximation for the applied field\cite{nakaneEvaluationNanoporesCandidates2002,rowghanianElectrophoreticCaptureDNA2013,rowghanianElectrophoresisDNACoil2013,grosbergDNACaptureNanopore2010,muthukumarTheoryCaptureRate2010,wanunuElectrostaticFocusingUnlabelled2010,nomidisDNACaptureClyA2018,qiaoVoltagedrivenTranslocationDefining2019} because it conserves the spherical symmetry and is quite accurate for distances much larger than pore radius $r_p$. The potential corresponding to the PCF can be written as
\begin{equation}
\label{eq:e_field_point}
U_e(r) = U_{PCF}(r)= \psi_o ~ \frac{r_e}{r},
\end{equation}
where $\psi_o= Q \Delta V$ is the drop in electrostatic energy of a particle of charge $Q$ when a voltage difference $\Delta V$ is applied across the system, and $r_e=r_p/({\pi}+\nicefrac{2\ell_p}{r_p})$ is the characteristic length  \cite{nomidisDNACaptureClyA2018,qiaoVoltagedrivenTranslocationDefining2019} of the potential outside a pore of radius $r_p$ and length $\ell_p$. The relevant electrophoretic charge is $Q=k_BT \mu /D$, with $\mu$ the electrophoretic mobility of the analyte. The field obtained from an exact solution of Laplace's equation\cite{farahpourChainDeformationTranslocation2013} will also be tested in our simulations for comparison. 

\begin{figure}[ht]
\begin{center}
\includegraphics[width=\columnwidth]{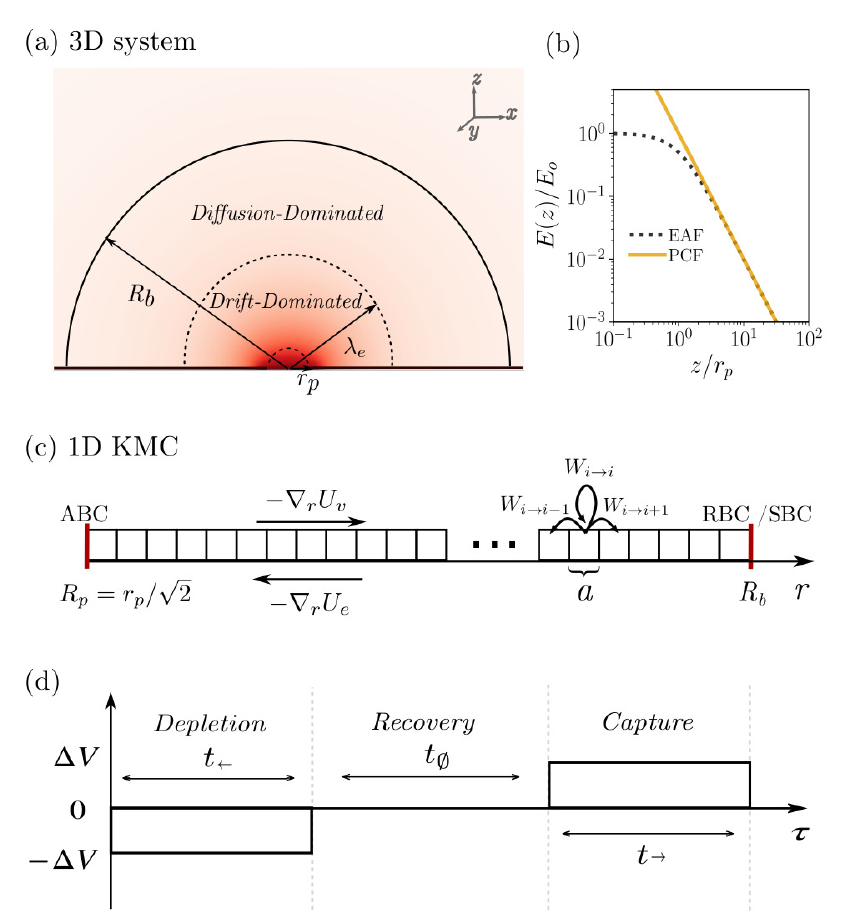}
\end{center}
\caption{(a) Schematic view of a nanopore system of size $R_b$:  The background color codes for the electric field strength (higher intensities near the pore). The dashed lines depict a hemisphere of radius $r_p$ (the pore size) and the capture radius $\lambda_e$. (b) Comparison between the exact axial (EAF) and point-charge (PCF) field approximations. The y-axis is scaled by the plateau value $E_o=E_{PCF}(r_p)$. (c) The 1D KMC model: $R_p=r_p/\sqrt{2}$ is the absorbing boundary; $R_b$ is the reflecting (RBC) or source (SBC) boundary; $-\nabla_r U_{v}$ is the entropic force; and $-\nabla_r U_{e}$ is the electric force. The jumping probabilities $W$ and lattice step size $a$ are also shown. (d) A schematic description of the pulse sequence. The field is applied in the reverse direction for a duration $t_{\shortleftarrow}$ in order empty a region near the nanopore; this is followed by an off period of duration $t_\emptyset$ during which the depleted region is partially refilled; finally, the capture period is of duration $t_\shortrightarrow$.}
\label{Fig_system}
\end{figure}

The capture radius $\lambda_e$ is generally defined as the radial distance at which the analyte's potential energy is equal to $k_BT$ (see Fig.~\ref{Fig_system}a); in our notation, it is given by
\begin{equation}
\label{eq:R*2}
 \lambda_e=\frac{\psi_o}{k_BT}~ r_e.
\end{equation} 
We use $\lambda_e$ as a measure of field intensity, the pore radius $r_p$ as the unit of length and $\tau_o \! = \! r_p^2/D$ as the unit of time.

The field-driven deterministic time\cite{qiaoVoltagedrivenTranslocationDefining2019} to drift from position $r_o$ to $r<r_o$ (using the PCF) is
\begin{equation}
\label{eq:tauEro2r}
    \tau_E(r_o,r)=\frac{ r_o^3-r^3}{ 3\lambda_e D}.
\end{equation}
For example, the \textit{cleanup time} $\tau_\lambda$ needed to empty the capture radius zone $\lambda_e$ is then
\begin{equation}
\label{eq:tauEL20}
    \tau_\lambda = \tau_E(\lambda_e,0)={\lambda_e^2}/{3 D} .
\end{equation}
The mean capture rate during the cleanup time $\tau_\lambda$ is 
\begin{equation}
    \label{eq:MeanRate}
    \overline{\rho}_\lambda \approx \tfrac{2 \pi}{3} \lambda_e^3 C_o/\tau_\lambda=2 \pi D \lambda_e C_o~.
\end{equation}

Although all of our results will be given in dimensionless units or as ratios, it is sometimes useful to be able to compare to actual experimental systems. In order to do that, we now look at two types of molecules: a short piece of ssDNA and a protein. Note however that this paper is not about either of these two cases: we use these numbers solely to establish the range of values that make sense for $\lambda_e$ (our simulations will use values between the two limits found below, namely $\lambda_e \approx 125 r_p$ and $\lambda_e \approx 22 r_p$).

For a 250 base ssDNA molecule\cite{nkodoDiffusionCoefficientDNA2001}, the relevant parameters are $D \! \approx \! 17\,\mu m^2/s$ and $\mu \approx 4.1 \times 10^{4}\,\mu m^2/Vs$, giving $Q \approx 60\, e$ ($\approx \nicefrac{1}{4}$ of the nominal charge). With a voltage $\Delta V \! = \! 400\,mV$, the potential energy is $\psi_o/k_BT \approx 900$.  For a pore of radius $r_p \! = \! 5\,nm $ and length $\ell_p\!=\!2\,r_p$,  we obtain $r_e=r_p/(4+\pi)$ and $\lambda_e \approx 125\, r_p \approx 1\,\mu m$. The basic time unit is then $\tau_o\!=\!r_p^2/D \approx 1.5 \,\mu sec$.

For a globular protein (for example Lysozyme\cite{allisonModelingElectrophoresisLysozyme1997}), the relevant parameters are $D \! \approx \! 100\,\mu m^2/s$, $\mu \approx 8 \times 10^{3}\,\mu m^2/Vs$, $Q \approx 10\, e$ and hydrodynamic radius $R_h\approxeq 2nm$. With a voltage $\Delta V \! = \! 400\,mV$, the potential energy is $\psi_o/k_BT \approx 160$.  For a pore of radius $r_p \! = \! 5\,nm $ and length $\ell_p\!=\!2\,r_p$,  we obtain $r_e=r_p/(4+\pi)$ and $\lambda_e \approx 22\, r_p$. The basic time unit is then $\tau_o\!=\!r_p^2/D \approx 0.25\,\mu sec.$

The stationary solution $\partial{C(r,t)}/\partial{t} \! = \! 0$ of eqs.~\ref{eq:diff} and~\ref{eq:e_field_point} with an absorbing boundary $C(R_p,t)=0$ in an infinite system with $C(r \shortrightarrow \infty,t)=C_o$ is given by \cite{grosbergDNACaptureNanopore2010,nomidisDNACaptureClyA2018} 
\begin{equation}
\label{eq:sol_diff_steady}
C(r)=C_o~\times~\frac{1-
\exp\left(-\lambda_e(1/R_p-1/r)\right)}{1-\exp(-\lambda_e/R_p)}.
\end{equation}
The location of the absorbing boundary $R_p$ is somewhat arbitrary since the pore has a finite width, unlike what eq.~\ref{eq:e_field_point} suggests. Previous papers used $R_p=r_p$, but this overestimates the surface area for capture since an hemisphere of radius $r_p$ has a surface area $2\pi r_p^2$ which is larger than that of the pore, $\pi r_p^2$. Instead, we use $R_p=r_p/\sqrt{2}$ to conserve the surface area.

The stationary concentration $C(r)$ rapidly increases from 0 at $r=R_p$ to $C_o$ since we generally have $\lambda_e \gg R_p$. A characteristic length can then be defined from eq.~\ref{eq:sol_diff_steady}: the depletion distance $r_d$ at which $C(r_d)/C_o=1-1/e$ is
\begin{equation}
    r_d \approx \frac{R_p}{1-{R_p}/{\lambda_e}},~~~~~\lambda_e \gg R_p.
\end{equation}
The width of the depletion region is thus of order $\sim R_p$. Using eq.~\ref{eq:tauEro2r}, the \textit{depletion time} $\tau_d=\tau_E(r_d,R_p)$ needed to establish the depletion zone $r_d$ is
\begin{equation}
\label{eq:tau_depeltion}
    \tau_d \approxeq \frac{R_p^4}{\lambda_e^2D }=\frac{1}{4} \left( \frac{r_p}{\lambda_e} \right)^2 \tau_o = \frac{3}{4} \left( \frac{r_p}{\lambda_e} \right)^4 \tau_\lambda.
\end{equation}
The depletion region is much smaller than $\lambda_e$, and its relaxation time is small compared to both $\tau_\lambda$
and $\tau_o$.

\section{Simulation methodology}
\label{section: KMC}

Simulating the capture process in $d=3$ dimensions would limit us to small systems and short times. In this section, we first show how to transform the 3D problem into a 1D one by replacing volume effects by an entropic force. We then design the corresponding lattice Kinetic Monte Carlo (KMC) algorithm for a point-like particle, and describe the boundary conditions. The mapping process is shown in Figs.~\ref{Fig_system}a and c.

\subsection{Fokker Planck Equation }
\label{section:FPE}

Building a KMC algorithm from a Fokker-Planck equation is simple due to the equivalence with the Master equation formulation. However, the drift-diffusion equation in spherical coordinates in $d$-dimensions
\begin{equation}
\label{eq:diff2}
\frac{\partial C(r,t)}{\partial t} \!  = \! \frac{D}{r^{d-1}} \frac{\partial }{\partial r} \! \left[ r^{d-1} \! \left( \! \frac{\partial C(r,t)}{\partial r} ~ + ~ \frac{C(r,t)}{k_BT} 
\frac{\partial U_e}{\partial r} \! \right) \! \right]
\end{equation}
is not strictly speaking a Fokker Planck equation\cite{n.g.vankampenStochasticProcessesPhysics2007}  for $d\neq1$ as it cannot be written as a conservation equation of the form
\begin{equation}
\frac{\partial C(r,t)}{\partial t} = -\frac{\partial }{\partial r}\left[ g(r;d)  C(r,t)  - D \frac{\partial C(r,t)}{\partial r}\right], \\
\end{equation}
where $g(r;d)$ would then be a drift term. This is due to the "centrifugal" term \cite{rednerGuideFirstpassageProcesses2001} $\frac{D(d-1)}{r} \frac{\partial }{\partial r} C(r,t)$ which appears when we expand the first term of the \textit{rhs} of eq~\ref{eq:diff2}.
Nonetheless, if use the radial concentration
\begin{equation}
 \label{eq:rho}
\widetilde{C}(r,t) = S(r;d) C(r,t),
\end{equation}
with $S(r;d)=\frac{2\pi^{d/2} }{\Gamma(d/2)}r^{d-1}$ the surface of a $d$-dimensional sphere, eq.~\ref{eq:diff2} directly reduces to\
\begin{equation}
 \label{eq:CL}
\frac{\partial \widetilde{C}(r,t)}{\partial t} = - \frac{\partial J(r,t)}{\partial r} ,
\end{equation}
where the radial flux $J(r,t)$ is given by
\begin{equation}
 \label{eq:PF}
\begin{aligned}
 J(r,t) =&~ -\underbrace{\frac{D}{k_BT}\widetilde{C}(r,t)\frac{\mathrm{d}U_v }{\mathrm{d}r} }_{J_v:~\mathrm{volume ~ drift}} -\underbrace{\frac{D}{k_BT}\widetilde{C}(r,t)\frac{\mathrm{d}U_e }{\mathrm{d}r}}_{J_e:~\mathrm{electrostatic ~ drift}} \\
 &~-\underbrace{D \frac{\partial \widetilde{C}(r,t)}{\partial r} }_{\mathrm{J_F:~Fick's ~ law}},
\end{aligned} 
\end{equation}
with $U_v(r) \! = \! (1-d)k_BT \ln(r)$.
We thus mapped a spherically symmetric "$d>1$" drift-diffusion problem onto a 1D process. The 1D projection greatly reduces the amount of memory needed to simulate large $d>1$ systems. This added a virtual entropic potential $U_v(r)$ and a "volume" drift $J_v$ pushing the particles away from $r=0$ due to the fact that there is more volume far from the origin. The electrostatic drift $J_e$ attracts the particles toward the center at $r=0$. Given eq.~\ref{eq:PF}, the stationary distribution $J(r,t)=0$ 
satisfies
\begin{equation}
 \label{eq:STA}
 \widetilde{C}^s(r) \propto \exp[-U(r)/k_BT],
 \end{equation}
where the effective potential energy is $U \! = \! U_v + U_e$.

\subsection{Kinetic Lattice Monte Carlo Algorithm}
\label{section:LMCA}

We consider a 1D system where particles can jump
between adjacent lattice sites $i$ and $i+1$ with probabilities $W_{i \leftrightarrow i+1}$ as shown in Fig.~\ref{Fig_system}c. The latter must satisfy detailed balance in order to insure microscopic reversibility:
\begin{equation}
\label{eq:DBE}
\widetilde{C}^s_i W_{i \shortrightarrow i+1} = \widetilde{C}^s_{i+1} W_{i+1 \shortrightarrow i}~.
\end{equation}
Using eq.~\ref{eq:STA}, we obtain
\begin{equation}
\label{eq:DBE2}
\frac{W_{i \shortrightarrow i+1}}{W_{i+1 \shortrightarrow i}} = \exp(\Delta \epsilon_i),
\end{equation}
with $\Delta \epsilon_i \! = \! \epsilon_{i+1} - \epsilon_i$ and $\epsilon_i \! = \! U_i/k_BT$. To link these parameters to local dynamics, we use the closure relation 
\begin{equation}
\label{eq:CR}
W_{i \shortrightarrow i+1}+W_{i+1 \shortrightarrow i} = 2D {\Delta t}/{a^2},
\end{equation}
where $a$ is the lattice step size and $\Delta t$ is the time step to be used for the simulations. The end result is
\begin{equation}
\label{Pp}
W_{i\shortrightarrow i+1} = \frac{2D}{a^2} \times \frac{1}{1 + \exp\left(\Delta \epsilon_i\right)} \times \Delta t,
\end{equation}
\begin{equation}
\label{Pm}
W_{i+1\shortrightarrow i} = \frac{2D}{a^2} \times \frac{1}{1 + \exp\left(-\Delta \epsilon_i\right)} \times \Delta t.
\end{equation}
Since the probability of not jumping during a time step, $W_{i\shortrightarrow i}=1-W_{i\shortrightarrow i-1}-W_{i\shortrightarrow i+1}$, must be $\ge 0$ $\forall i$, we have \begin{equation} 
\label{eq_KMC_condi}
\Delta t \leq 1/\max[R_{i \shortrightarrow i-1}+R_{i \shortrightarrow i+1}],
\end{equation}
where the hopping rates are $R_{i \shortrightarrow i\pm1}= W_{i \shortrightarrow i\pm1}/ \Delta t$. In order to achieve optimal accuracy, we use a time step $\Delta t = 1/3\,\Delta t_{max}$ in our simulations to insure $W_{i\rightarrow i}\approx \tfrac{2}{3}$ everywhere on the lattice, so that the local bias does not affect the diffusion coefficient of the particle\cite{chubynskyOptimizingAccuracyLattice2012,gauthierBuildingReliableLattice2004,dehaanImportanceIntroducingWaiting2011}.

This KMC algorithm can be used in two different ways:

I) To study the motion of a single particle: as usual, a random number is then generated at each time step to select the next move that will be attempted. 

II) To follow a population of particles: The time evolution of $\tilde{C}$ can be studied by iterating the master equation. The concentration $\tilde{C}_{i}^{j+1}$ at lattice $i$ and time step $j\!+\!1$ reads
\begin{equation}
 \label{eq:master}
{\tilde{C}_{i}^{j+1}}=W_{i-1\shortrightarrow i}\tilde{C}_{i-1}^{j}+W_{i+1\shortrightarrow i}\tilde{C}_{i+1}^{j}+W_{i\shortrightarrow i}\tilde{C}_{i}^{j},   
\end{equation}
where $i,j$ are integers.

\subsection{Boundary conditions}
We use three different types of boundary conditions, as shown in Fig.~\ref{Fig_system}c:\\

Absorbing Boundary Conditions to model the capture by the nanopore (\textbf{ABC}): We consider that the ABC is in the center of the lattice site at a distance $R_p$ from the origin; the boundary condition then reads $C(p-1)=0$, where $p=R_p/a$. The corresponding master equation is 
\begin{equation}
{\tilde{C}_{p}^{j+1}}=W_{p+1\shortrightarrow p}\tilde{C}_{p+1}^{j}+W_{p\shortrightarrow p}\tilde{C}_{p}^{j}.  
\end{equation}

Reflecting Boundary Conditions (\textbf{RBC}): In one series of simulations, the walls of the cavity of size $R_b$ are replaced by a RBC placed in the center of the last lattice site, $b=R_{b}/a$; jumps from $i=b$ to $b+1$ are rejected (there is no particle flux across the boundary). The corresponding master equation is 
\begin{equation}
{\tilde{C}_{b}^{j+1}} \! = \! W_{b-1\shortrightarrow b}\tilde{C}_{b-1}^{j}+W_{b\shortrightarrow b}\tilde{C}_{b}^{j}+W_{b\shortrightarrow b+1}\tilde{C}_{b}^{j}. 
\end{equation}

Source Boundary Conditions (\textbf{SBC}):
In some simulations, the cavity walls are replaced by a source that mimics an infinite system (or reservoir) at fixed concentration $\tilde{C}(j \ge b)=\tilde{C}_o$. The corresponding master equation is 
\begin{equation}
{\tilde{C}_{b}^{j+1}}=W_{b-1\shortrightarrow b}\tilde{C}_{b-1}^{j}+W_{b\shortrightarrow b}\tilde{C}_{b}^{j}+W_{b+1\shortrightarrow b}\tilde{C}_{o}.   
\end{equation}
We use a lattice step size $a=R_p/10$ for the simulations in Sections 4.1 and 4.2, and $a=r_p/100$ for the rest of the paper. The simulation time required to achieve steady-state depends on system size, the level of discretization (lattice step size and choice of time step) and field intensity. Let us illustrate this using Test 1 presented in Section 4.1 as an example: the simulations are carried out using Python 3.7 with NumPy 1.19; the typical simulation time for a system described in Section 4.1 is $\approx 30$ minutes for $10^5$ iterations on a single core processor (2.6 GHz Intel Core i7), corresponding to $t=636\,\tau_t$. The figures are plotted using the Matplotlib 3.2.2 package.\\

\subsection{The electric field}

We use two different electric field approximations in our simulations: The spherically symmetric field from the point-charge field (PCF) approximation is given by  
\begin{equation}
\label{seq:EqpiontE}
{E_{PCF}}(r)= -\Delta V ~ \frac{r_e}{ r^2}.
\end{equation}
As discussed previously\cite{qiaoVoltagedrivenTranslocationDefining2019,kowalczykModelingConductanceDNA2011,farahpourChainDeformationTranslocation2013}, the actual electric field is identical to the PCF at large distance ($r>2\,r_p$) but flat and not spherically symmetric near the pore (the differences are basically found at small polar angles $\theta$ when $r<2\,r_p$). Our 1D model is a projection of a spherically symmetric 3D capture system. In order to investigate the impact of the flat field in our 1D KMC simulations, we neglect the small angular dependence of the field near the pore and use the exact field\cite{farahpourChainDeformationTranslocation2013} in the vertical direction right above the pore ($\hat{z}$ in Fig.~\ref{Fig_system}b). We call this the Exact Axial Field (EAF) approximation. In 1D simulations, the  electric potential at distance $r$ then reads
\begin{equation}
 \label{seq:sol}
V_{EAF}(r) = \tfrac{r_e }{r_p} ~ \Delta V \arctan \left( \tfrac{r}{r_p} \right),
\end{equation}
and the corresponding electric field is
\begin{equation}
 \label{seq:Electricfield}
  {E_{EAF}}(r) = -\Delta V ~ \frac{r_e}{r_p^2+ r^2}.
\end{equation}
Note that ${E_{EAF}}(r \! \gg \! r_p) \approx E_{PCF}(r)$, in agreement with the PCF approximation, eq.~\ref{seq:EqpiontE}. The field plateaus at ${E}_{o}= E_{PCF}(r_p)$ close to the pore, as shown in Fig.~\ref{Fig_system}b.

\section{Results}
\label{section:results}

We first test our KMC algorithm for diffusion-limited (no field) absorption by a sphere, both in finite and infinite systems. We then add the electric field to study capture by the nanopore, again in both finite and infinite systems, and simulation results with the PCF and AEF field approximations are compared. We also examine how a particle escapes from the nanopore when the polarity of the electric field is reversed.

\subsection{Test 1: Diffusion-limited absorption by a sphere  with a source boundary}

\label{sec:diffusiontest1}

The solution of the diffusion equation with $U_e=0$ (no external field), an absorbing sphere of radius $R_p$ and the boundary conditions $C(R_p,t)=0$ and $C(\infty,t)=C_o$ is\cite{rednerGuideFirstpassageProcesses2001,bressloffStochasticProcessesCell2014}
\begin{equation}
\label{eq:diff_lim_solution}
    C(r,t)=C_o\left(1-\tfrac{R_p}{r}\right)+\tfrac{R_pC_o}{r}\erf{\left[\tfrac{r-R_p}{\sqrt{4Dt}}\right].}
\end{equation}
The last term is negligible in the steady state, $t\shortrightarrow\infty$; the result is then identical to the $\lambda_e \shortrightarrow 0$ limit of eq.~\ref{eq:sol_diff_steady}, as it should. The time dependent capture rate is
\begin{equation}
\label{eq:diff_lim_R}
    \rho(R_p,t) \! = \! 4\pi R_p^2 D \left.{\tfrac{\partial C}{\partial r}}\right\rvert_{R_p} \! = \! \rho_{s}^o  \left(1+\tfrac{R_p}{\sqrt{\pi Dt}}\right) \! .
\end{equation}
The capture rate decays to the steady state value $\rho_{s}^o \! = \! \rho(R_p,\infty) \! = \! 4\pi DR_pC_o$ with a relaxation time $\tau_t \! = \! R_p^2/D\pi$. Interestingly, if we replace $R_p$ by $\lambda_e$ in the expression for $\rho_s^o$, we recover the capture rate for a nanopore, eq.~\ref{eq:MeanRate} (to within a factor of 2 because eq.~\ref{eq:MeanRate} is for a half-space); the field thus pushes the capture radius from $R_p$ to $\lambda_e$.   

\begin{figure}[ht]
\begin{center}
\includegraphics[width=\columnwidth]{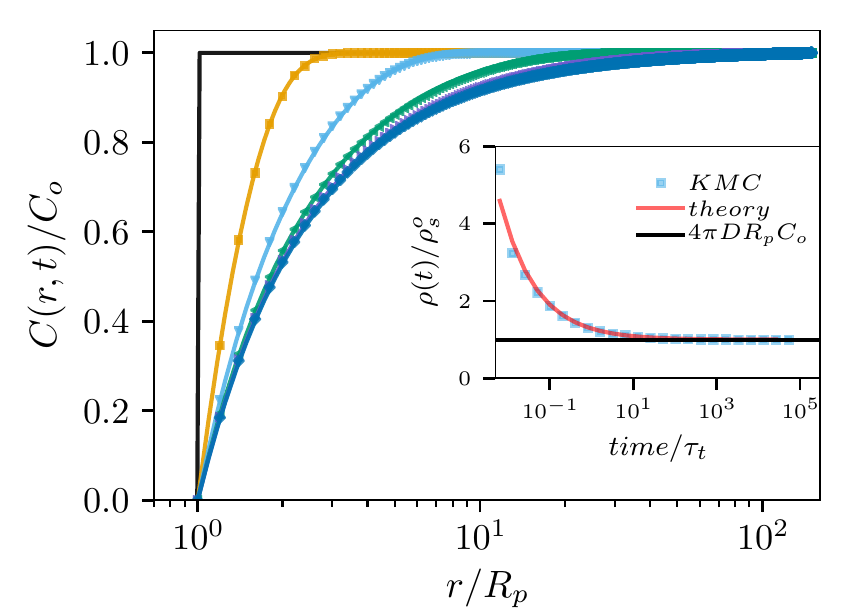}
\end{center}
\caption{Scaled concentration $C(r,t)/C_o$ \textit{vs} $r/R_p$ at different times $t$ (from left to right: $t=0,\textcolor{orange}{0.1},\textcolor{Cerulean}{2},\textcolor{ForestGreen}{26},\textcolor{BlueViolet}{420}$ and $\textcolor{NavyBlue}{6700}\,\tau_t$) for diffusion-limited absorption by a sphere of radius $R_p$ with a source boundary at $R_b=150 R_p$. The data points are from simulations and the solid lines are from the exact solution, eq.~\ref{eq:diff_lim_solution}. Inset: Scaled capture rate $\rho(t)/\rho_s^o$ \textit{vs} time; the data points (\textcolor{SkyBlue}{$\blacksquare$}) are from simulations, the red line is from eq.~\ref{eq:diff_lim_R} and the horizontal line is the predicted steady-state.}
\label{Fig:diff_C}
\end{figure}

The simulation system of size $R_b=150~R_p$ has a uniform loading $C(r,0) \! = \! C_o$ for $R_p \! < \! r \! < \! R_b$, an absorbing boundary $C(R_p,t) \! = \! 0$ and a source boundary $C(R_b,t) \! = \! C_o$  (in order to mimic an infinite system).  Fig.~\ref{Fig:diff_C} shows the radial dependence of $C(r,t)$ at different times $t$ while the inset shows the time-dependent capture rate: the results agree with theory. 

\subsection{Test 2: Diffusion-limited absorption by a sphere in the presence of an outer reflecting boundary}
\label{sec:diffusiontest2}

We now replace the SBC at $R_b$ by a reflecting boundary (the total number of particles thus decreases with time), and we compare our results to those of Section \ref{sec:diffusiontest1}, still in absence of an external field. Here, the concentration is given by (Appendix \ref{sec:rbc_test})
\begin{equation}
\label{eq:diff_C_rbc}
C(r,t) = C_o \sum_{n=1}^\infty B_n \frac{\sin(k_n(r-R_p))}{r/R_p}\exp(-t D k_n^2)
\end{equation} 
where the wavenumbers $k_n$ are the roots of
\begin{equation}
\label{eq:diffmax_rob}
\tan \left( k_n (R_b - R_p) \right) = R_b k_n
\end{equation}
and the weight of the $n^{th}$ decay mode is given by
\begin{align}
\label{eq:diffmax_Bn}
B_n &= \frac{\int_0^{R_b \! - \! R_p}(r+R_p) \sin(\lambda_n r) \mathrm{d}r}{R_p\int_0^{R_b-R_p} \sin^2(\lambda_n r) \mathrm{d}r} \nonumber  \\
& = \tfrac{\sin(k_n (R_b \! - \! R_p))-k_n R_b \cos(k_n(R_b \! - \! R_p))+k_n R_p}{R_p k_n^2 \left( \frac{R_b-R_p}{2} - \frac{\sin(2k_n(R_b \! - \! R_p))}{4k_n}\right)}.
\end{align}
The time dependent capture rate is
\begin{equation}
\label{eq:diffmax_rho}
\rho(R_p,t) = 4\pi D C_o R_p\sum_{n=1}^\infty B_n k_n R_p  \exp(-t D k_n^2).
\end{equation} 
At long times $t\gg R_b^2/D$, the concentration reduces to
\begin{equation}
\label{eq:diff_C_rbc_lt}
\frac{C(r,t)}{C_o} \approx  B_1 \frac{\sin(k_1(r-R_p))}{r/R_p}\exp(-t D k_1^2),
\end{equation}
while the capture rate decays as
\begin{equation}
\label{eq:diffmax_rholim}
\rho(R_p,t\rightarrow\infty) \approx \rho_s^o \times R_p  k_1 B_1 \exp(-t D k_1^2),
\end{equation} 
where $k_1$ and $B_1$ are the wavenumber and weight of the longest mode, respectively. The final decay time is thus 
\begin{equation}
\label{eq:tau_c}
\tau_1={1}/{Dk_1^2}.
\end{equation}
In the large box limit $R_b \! \gg \! R_p$, we obtain $k_1R_b \approx 1.571$, $\tau_1 \! \approx \! 0.405~R_b^2/D$ and $B_1 \! \approx \! 0.811~R_b/R_p$. Equation~\ref{eq:diffmax_rholim} then predicts that $\rho(t)$ becomes smaller than the infinite system plateau rate $\rho_s^o$ when $t > \tau_c \approx 0.2416~ \tau_1$. In other words, $\tau_1 \sim R_b^2/D$ is the time required to measurably deplete the box as a whole.

\begin{figure}[ht]
\begin{center}
\includegraphics[width=\columnwidth]{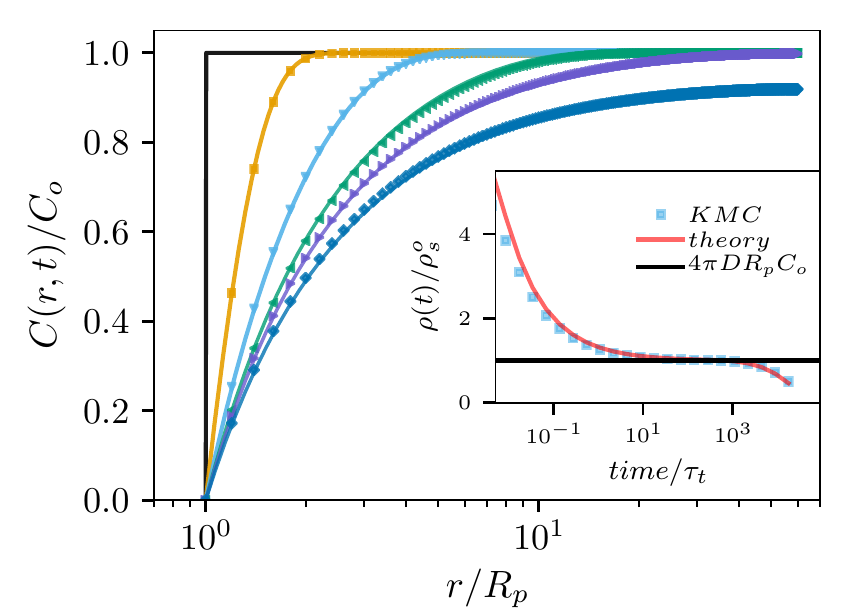}
\end{center}
\caption{Scaled concentration $C(r,t)/C_o$ \textit{vs} $r/R_p$ at different times $t$ ($=0,\textcolor{orange}{0.03},\textcolor{Cerulean}{0.5},\textcolor{ForestGreen}{9},\textcolor{BlueViolet}{140}$ and $\textcolor{NavyBlue}{2300}\,\tau_t$) for diffusion-limited absorption by a sphere of radius $R_p$ with a reflecting boundary at $R_b=60 R_p$. The data points are from simulations while the solid lines show the analytical solution, eq.~\ref{eq:diff_C_rbc}. Inset: Scaled capture rate \textit{vs} time; the data points (\textcolor{SkyBlue}{$\blacksquare$}) are from simulations, the red line is from eq.~\ref{eq:diffmax_rho} and the horizontal line is the steady-state value for an infinite system.}
\label{Fig:diff_test2}
\end{figure}

We use the simulation setup of Section \ref{sec:diffusiontest1} but replace the source boundary by a reflecting one (at $R_b=60 R_p$). Figure~\ref{Fig:diff_test2} shows the radial dependence of $C(r,t)$ at different times while the inset gives the time-dependent capture rate: the data agree with the theory. The rough estimate above gives a critical time $\tau_c \approx 10^3\tau_t$ here, in agreement with the inset data.

\subsection{Field-driven capture rate with a source boundary}
\label{sec:sbc_field}

\begin{figure}[ht]
\begin{center}
\includegraphics[width=\columnwidth]{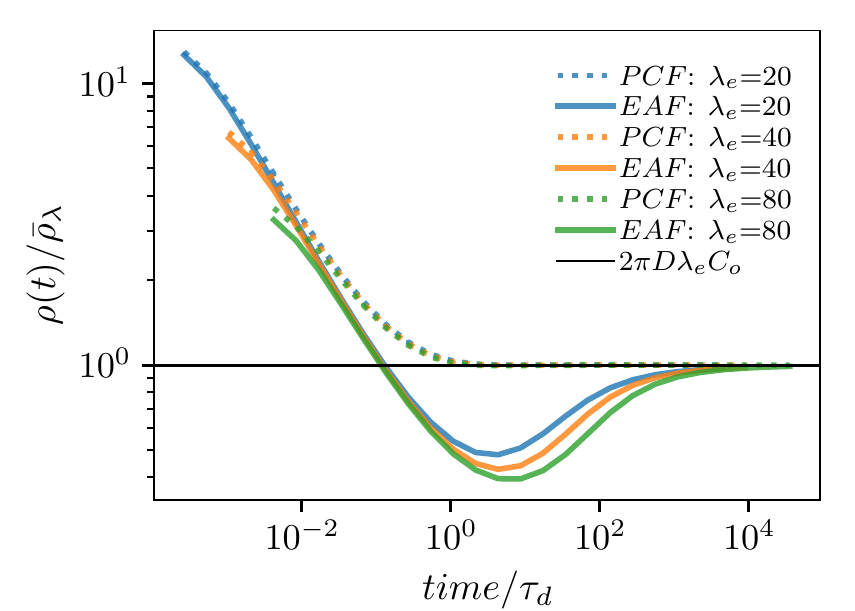}
\end{center}
\caption{Capture rate $\rho(t)$ \textit{vs}
 time for field intensities $\lambda_e=20$, $40$ and $80\,r_p$. The pore (absorbing) boundary is located at $R_p=r_p/\sqrt{2}$ while the source boundary is at $R_b=2\lambda_e$. The horizontal line is the steady state capture rate predicted by eq.~\ref{eq:MeanRate}. The dotted lines are from simulations done using the PCF, eq.~\ref{eq:e_field_point}, while the solid lines give the simulation results when the EAF is used, eq.~\ref{seq:Electricfield}. The time axis is rescaled by the steady-state time $\tau_d(\lambda_e)$ while the capture rates are rescaled by the steady-state theoretical value $\overline{\rho}_\lambda=2 \pi DC_o \lambda_e $.}
\label{Fig:sbc_flux_Nlambdae}
\end{figure}

\begin{figure}[ht]
\begin{center}
\includegraphics[width=\columnwidth]{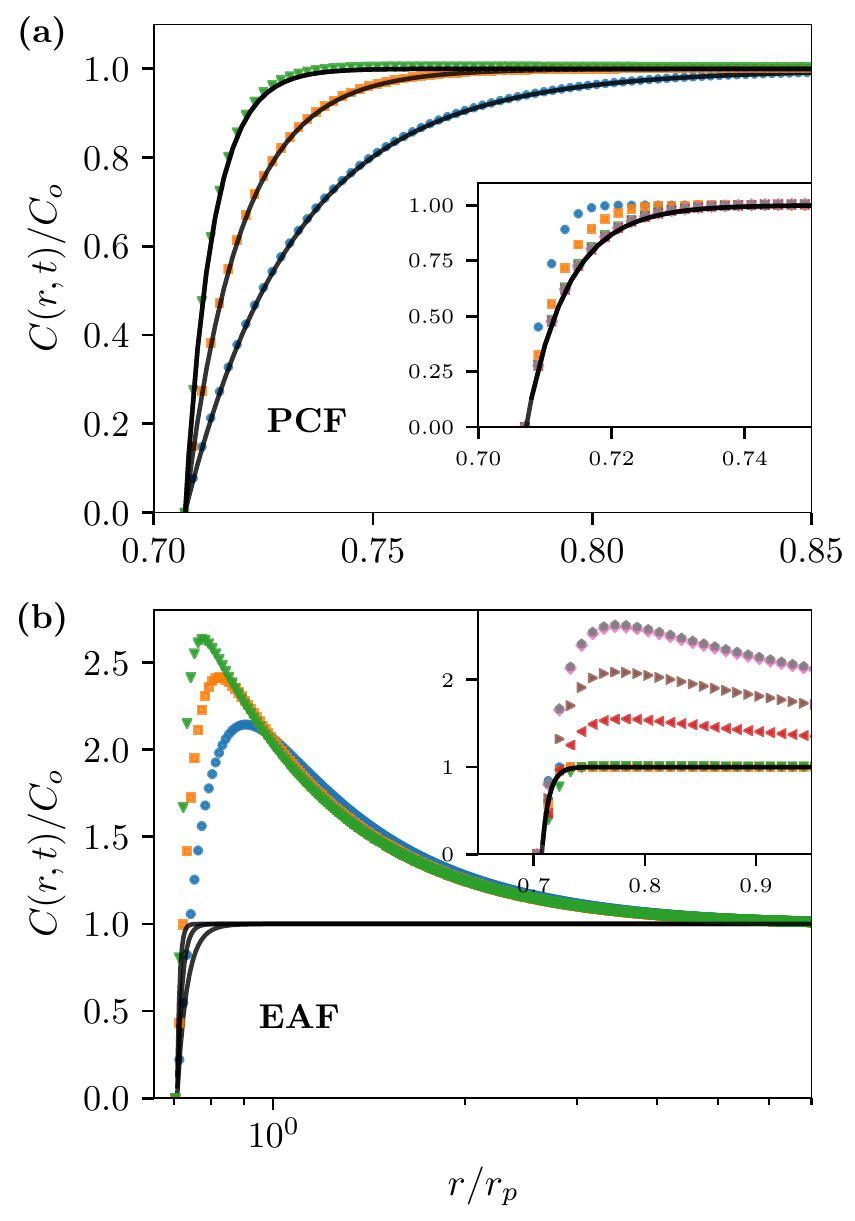}
\end{center}
\caption{Scaled concentration $C(r,t)/C_o$ \textit{vs} $r/r_p$ for field intensities $\lambda_e \! = \! 20$~(\textcolor{NavyBlue}{\Large{$\bullet$}}), $40~$(\textcolor{Orange}{$\blacksquare$}) and $80~r_p$~(\textcolor{OliveGreen}{\large{$\blacktriangledown$}}). The pore is located at $R_p=r_p/\sqrt{2}$ while the source boundary is at $R_b=2\lambda_e$. (a): Simulation results at time $t=10 \tau_d$ obtained using the PCF. (b): Simulation results at time $t=10000 \tau_d$ obtained using the EAF. In both cases, the solid lines show the PCF prediction, eq.~\ref{eq:sol_diff_steady}. Insets: Simulation data at times $0.07, 0.3, 1, 70, 300, 9000$ and $20000\tau_d$, with $\lambda_e=80\,r_p$.}
\label{Fig:sbc_Nlambda_C}
\end{figure}

In this section, we use a source boundary at $r \! = \! R_b \! = \! 2 \lambda_e$ and an absorbing one at $R_p=r_p/\sqrt{2}$ (the nanopore). Figure~\ref{Fig:sbc_flux_Nlambdae} shows the time dependence of the capture rate $\rho(t)$ for different field intensities $\lambda_e$, for both the PCF and EAF approximations. The curves collapse remarkably well when the time axis is rescaled by the theoretical PCF transient time $\tau_d(\lambda_e)$. The PCF capture rate rapidly decays to the predicted value at roughly $t \approx \tau_d$. When we use the EAF instead, the curves overlap up to $t \approx \tau_d$, followed by a deep undershoot, and finally the same final rate is reached for times $t \gtrsim 10^4\tau_d$. The fact that the EAF field is lower than the PCF field near the pore has two effects: 1) the time required to reach the steady-state is increased; 2) the width $r_d$ of the region where $C(r)$ is not flat is broader (see Fig.~\ref{Fig:sbc_Nlambda_C}b for example). Both of these effects contribute to the large increase in the transient time.

Figure~\ref{Fig:sbc_Nlambda_C} shows several steady-state concentration profiles $C(r,t \gg \tau_d)$ for both PCF and EAF. The $C(r,t \gg \tau_d)$  PCF curves, Fig.~\ref{Fig:sbc_Nlambda_C}a,  agree perfectly with the analytical solution,  eq.~\ref{eq:sol_diff_steady}; in particular, the depletion zones are barely larger than $R_p$ and get narrower at higher field. However, in the EAF case (Fig.~\ref{Fig:sbc_Nlambda_C}b), there is a peak near the pore due to the locally flat field, and its position shifts closer to the pore when the field increases. Time-dependent concentration profiles are shown in the insets for $\lambda_e=80\,r_p$. A depletion zone quickly forms in the PCF case, as expected. However, the EAF concentration increases near pore due to the slower capture rate caused by the flat field; the increasing local concentration results in a higher capture rate; for times $t \gtrsim 10^4 \tau_d$, however, these two effects balance each other and the steady-state is reached.    

\subsection{{Field-Driven capture rate with a reflecting boundary}}

\begin{figure}[ht]
\begin{center}
\includegraphics[width=\columnwidth]{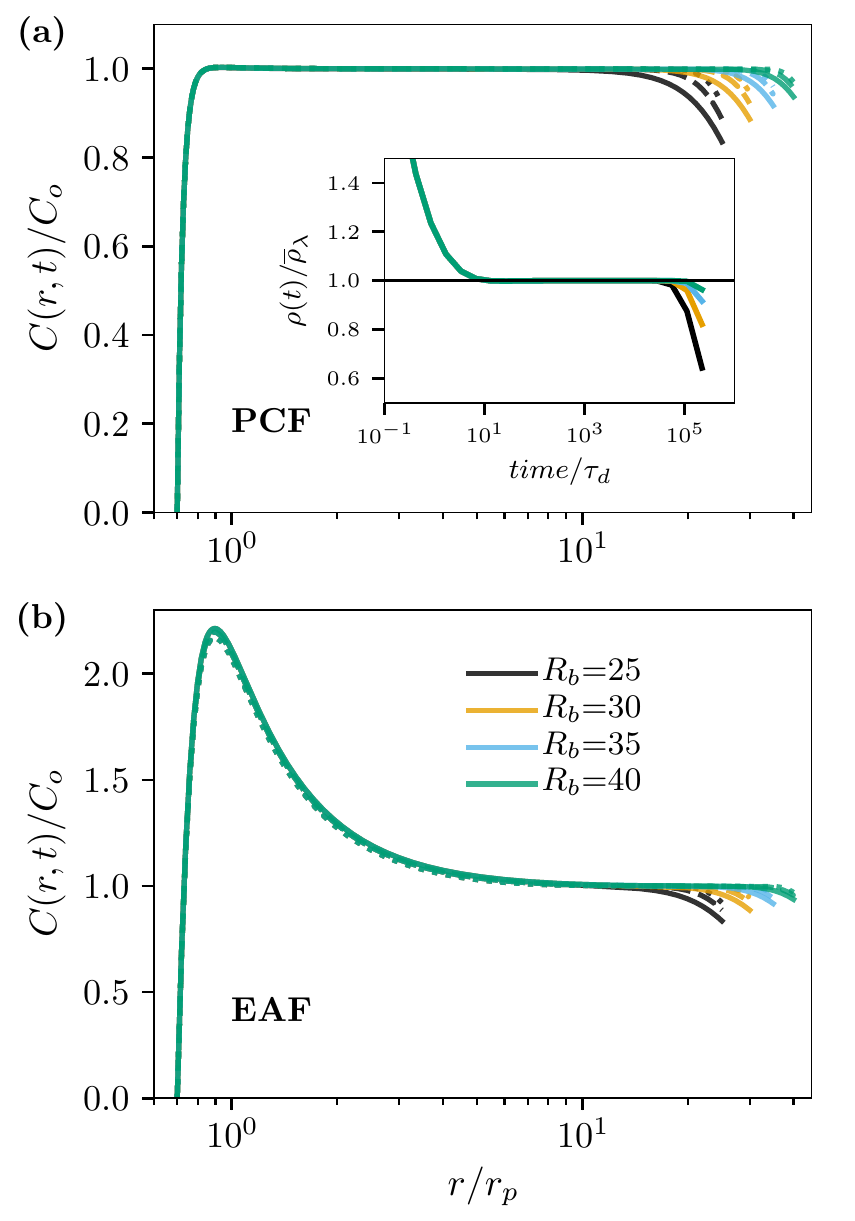}
\end{center}
\caption{Scaled concentration $C(r,t)/C_{o}$ \textit{vs} $r/r_p$  at times $t=1700~(\cdots), 3500~(\mbox{-}~\mbox{-}~\mbox{-})$ and $7000~(\textbf{\mbox{---}})\,\tau_d$ for a field intensity $\lambda_e=20\,r_p$. The systems have a reflecting boundary at $r=R_b$, as indicated, and the pore is located at $R_p=r_p/\sqrt{2}$. (a) Simulation results obtained using the PCF, eq.~\ref{eq:e_field_point}. (b): Simulation results obtained using the EAF, eq.~\ref{seq:Electricfield}. Inset in (a): Capture rate $\rho(t)$ \textit{vs} time. The horizontal line is the steady state capture rate predicted by eq.~\ref{eq:MeanRate}.}
\label{Fig:rbc_NRb_C}
\end{figure}

We now replace the SBC used in Section \ref{sec:sbc_field} with an RBC to investigate the impact of finite system size on capture. Figure~\ref{Fig:rbc_NRb_C} shows the concentration profile at three different times for a field intensity $\lambda_e=20\,r_p$ and different box sizes $R_b > \lambda_e$. We observe similar long-time behavior for both the PCF and EAF fields: the concentration near the outer, reflecting wall decreases with time since no new particles arrive from infinity in this case. Unsurprisingly, these effects happen earlier and are more severe for smaller system sizes $R_b$. 

The rate at which the outer depletion zone propagates inward can be estimated as follows. If we assume that the concentration profile is a step function with $C \approx C_o$ up to the beginning of the depletion zone and zero beyond, the equation for the location $r(t)$ of the front is simply 
\begin{equation}
     2 \pi r(t)^2 C_o dr=-\rho_s^o dt=-2 \pi D C_o \lambda_e dt,
\end{equation} 
with $r(0)=R_b$. The solution is
\begin{equation}
    r(t)= R_b \times \left(1-{3 \lambda_e D t}/{R_b^3} \right)^{1/3}.
\end{equation}
The time taken by this second depletion region to reach the pore is thus $\tau_b \approx R_b^3/3\lambda_e D=\tau_E(R_b,0)$. However, we expect that the capture rate will start being affected roughly when the depletion region reaches the capture radius at $r=\lambda_e$, i.e. at time $\tau_\rho \approx [1-(\nicefrac{\lambda_e}{R_b})^3]\tau_b$. The inset of Fig.~\ref{Fig:rbc_NRb_C}a shows the time dependence of the capture rates for the PCF case. The size of the box plays no role at short times, but $\rho(t)$ decreases at longer times -- similar to the field-free results in Fig.~\ref{Fig:diff_test2}. The rough theory described above overestimates the time at which this happens by a factor of 10, not surprising given the fact that the propagating front is not a step function and thus propagates faster than assumed here.

\begin{figure}
\begin{center}
\includegraphics{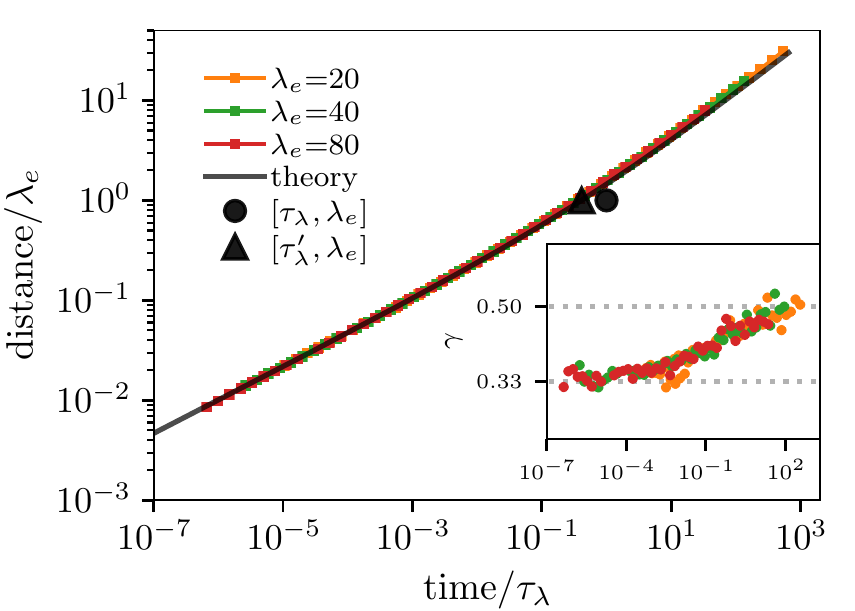}
\end{center}
\caption{Log-log plot of the mean distance migrated (in units of $\lambda_e$) at time $t$ (in units of $\tau_\lambda$) when the PCF polarity is reversed. The black line is from eq.~\ref{eq:new_tau_theory}. The filled circle ({\large{$\bullet$}}) shows the location of the $[\tau_\lambda; \lambda_e]$ point, while the triangle ({{$\blacktriangle$}}) uses the improved estimate of the clean up time $\tau_\lambda^{\prime}$ given by eq.~\ref{eq:new_tau_theory}. The particles are initially placed one lattice site above the pore and the data points ($r\geq R_p$) are averaged over 1000 trajectories. Inset: Local slope \textit{vs.} time. The expected values of $\nicefrac{1}{2}$ and $\nicefrac{1}{3}$ are marked by horizontal lines.}
\label{Fig:Rev}
\end{figure}

\subsection{Time reversal and particle escape}
\label{section:timereversal}
To better understand the dynamics of the particles at different radial distances $r$, we now study the inverse of the capture process by using an open boundary condition and an inverted PCF polarity. Our goal is to examine the transition between field-driven and diffusion-driven dynamics when a particle moves across the capture radius (the same transition occurs for escape and capture simulations, but the former are computationally more efficient).

We start the particles one lattice site above the pore and let them move away; however, we only record the data once the particles have reached the location $r=R_p$ of the absorbing boundary used in the previous section. The time dependence of the mean radial displacement $\overline{r}(t)$ can inform us about the relative importance of diffusion and electric drift during capture. Indeed, we expect that a particle moving away from the pore will go through two main phases: field-driven (as described by eq.~\ref{eq:tauEro2r}) for short distances, and then diffusive when $r>\lambda_e$ (and similarly, but in reverse order, for capture).

Figure~\ref{Fig:Rev} shows $\overline{r}(t)$ \textit{vs} $t$ for several field intensities $\lambda_e$, and the inset shows the local slope (\textit{i.e.}, the exponent $\gamma$ if $\overline{r}(t) \sim t^\gamma$). We clearly have two regimes, with $\gamma=\nicefrac{1}{2}$ (diffusion) at long times $t > \tau_\lambda$, preceded by $\gamma=\nicefrac{1}{3}$ (field-driven motion). 

The data for different field intensities collapse if the distances and times are rescaled by the capture radius $\lambda_e$ and the cleanup time $\tau_\lambda$, respectively, in agreement with our previous paper\cite{qiaoVoltagedrivenTranslocationDefining2019}. However, the curves do not quite go through the $[\tau_\lambda,\lambda_e]$ point as one might have expected. The reason for this is the fact that eq.~\ref{eq:tauEL20} considers only the deterministic effect of the field and neglects both diffusion and entropic effects. If we take volume (entropic) effects into account, the deterministic time $\tau_r{(r)}$ to reach a distance $r$ from the origin can be calculated by integrating the combined effects of the electric force $\nabla U_e$ and the entropic force $-\nabla U_v$:
\begin{equation}
\label{eq:new_tau_theory}
    \begin{aligned}
     \tau_r(r) &=\frac{k_BT}{D}\int_{0}^{r} \frac{\mathrm{d}r^{\prime}}{-\nabla U_v(r^{\prime})+\nabla U_e(r^{\prime})}\\
     &=\frac{\lambda_e^2\ln(2r/\lambda_e+1)-2r(\lambda_e-r)}{8D}.
\end{aligned}
\end{equation}
A more accurate estimate of the time to reach the radial distance $r=\lambda_e$ is thus $\tau_\lambda^{\prime}=\tau_r(\lambda_e)=\frac{3\ln(3)}{8}\tau_\lambda$. Figure~\ref{Fig:Rev} shows that $\tau_\lambda^{\prime}$ agrees nicely with the simulation results. In the limit $r \gg \lambda_e$ where diffusion dominates eq.~\ref{eq:new_tau_theory} gives $\tau_r \approx {r^2}/{4D}$ instead of the expected $\tau_r \approx {r^2}/{6D}$ in three dimensions: this is the reason why the black curve in Fig.~\ref{Fig:Rev} is then below the simulation data. The EAF data are similar although not identical at short times due to the flat field  (not shown).

\section{Can we concentrate analytes using a time-varying field?}
\label{section:separation}

Our KMC algorithm allows us to also study molecular mixtures as well as time-dependent fields. As an example, we now investigate (as a proof of concept) a process by which it might be possible to favour the capture of one molecular species in a mixture of two types of molecules. 

The steady-state capture rate is given by eq.~\ref{eq:MeanRate}. Returning to the original variables, this can be written as $\rho=2 \pi C_o \Delta V r_e \mu$, showing that while the mobility $\mu$ affects the capture rate, the diffusion coefficient $D$ does not. Therefore, the most challenging situation would be to manipulate the capture rates of two molecules that happen to share the same mobility $\mu$. We will be using this hypothetical case for our proof-of-concept analysis.

We thus consider a mixture with two types of analytes having the same mobility $\mu_1=\mu_2$ but different diffusion coefficients, with $D_1<D_2$. The question we are asking ourselves here is simply: can we design a capture process that would favor one species?

In order to bias the translocation process, we have to exploit the fact that $D_1 \ne D_2$, and this implies that we periodically turn the field off so that diffusion can play a role. One approach is to briefly reverse the field polarity to create a depletion region near the pore, and then turn the field off to let the molecules fill this region by diffusion -- a slow process that will bring more of the fast-diffusing, smaller type 1 molecule in the vicinity of the pore. If we then apply the translocating field for a short period of time, the nanopore will capture more type 1 than type 2 molecules. Repeating this pulse sequence (Fig.~\ref{Fig_system}d) will lead to a capture process that is biased in favour of type 1 molecules. 

The duration of the three pulses in the sequence will be denoted $t_\emptyset$,  $t_{\shortleftarrow}$ and $t_\shortrightarrow$~. These pulse durations must be selected properly, as we now demonstrate. 

According to eq.~\ref{eq:tauEro2r}, the radius of the depletion region created during the reverse field phase is
\begin{equation}
\Delta r_{\shortleftarrow} \approx \sqrt[3]{3 \mu \Delta V r_e t_{\shortleftarrow}}~.
\end{equation}
During the recovery phase of duration $t_\emptyset$, the field is turned off and the molecules refill the depletion zone by diffusion. The change in radius of this region is thus
\begin{equation}
    \Delta r_\emptyset^{(1,2)}  \approx \sqrt{6D_{1,2} t_\emptyset}.
\end{equation}
Since $D_2 \! < \! D_1$, choosing $t_\emptyset$ such that $\Delta r_\emptyset^{(2)} \! < \! \Delta r_{\shortleftarrow}$ would minimize the number of type 2 molecules near the pore, while a $t_\emptyset$ that gives $\Delta r_\emptyset^{(1)} \ge \Delta r_{\shortleftarrow}$ would maximize capture of type 1 molecules. In other words, we need
\begin{equation}
    \label{eq:backpulse}
    \Delta r_\emptyset^{(1)} ~\ge ~\Delta r_{\shortleftarrow}~ >~ \Delta r_\emptyset^{(2)}.
\end{equation}
In the final phase, we want to capture the molecules in the region not yet fully refilled by the slowest type 2 molecules. Since this region has a radius $\Delta r^{(2)} \approx \Delta r_{\shortleftarrow} - \Delta r_\emptyset^{(2)}$, this means
\begin{equation}
\label{eq:capturetime}
    t_\shortrightarrow < \tau_E(\Delta r^{(2)},0)
\end{equation}
where $\tau_E$ can be estimated using  eq.~\ref{eq:tauEro2r}. 
One last condition must be satisfied since the type 1 molecules that reach the mouth of the nanopore must also have time to translocate across the entire nanochannel during the capture pulse of duration $t_\shortrightarrow$ (we assume that those who do are taken away immediately and permanently, e.g. by a liquid flow parallel to the wall). The electric field near nanopore ($r\leq r_p$) is flat with a field intensity of $E_o=E_{PCF}(r_p)=-\Delta V{r_e}/{r_p^2}$, which is essentially the same as the electric field inside the nanochannel. The distance travelled by the particles (both types) located inside the nanochannel (nc) during the forward field period is thus
\begin{equation}
\label{eq:drift_pore}
    r_{nc}=\mu E_o t_\shortrightarrow=\frac{\lambda_e D}{r_p^2}~t_\shortrightarrow
\end{equation}
The last condition is simply
\begin{equation}
\label{eq:min_translocation}
    r_{nc} > \ell_{nc}~,
\end{equation}
where $\ell_{nc}=\ell_p+2R_p$ is basically the effective length of the channel here. The three pulse durations must be chosen to satisfy eqs \ref{eq:backpulse}, \ref{eq:drift_pore} and \ref{eq:min_translocation} simultaneously. As usual, the performance of such a system will be a trade-off between high selectivity and high capture rates. We give an example below.

\begin{figure}[ht]
\begin{center}
\includegraphics[width=\columnwidth]{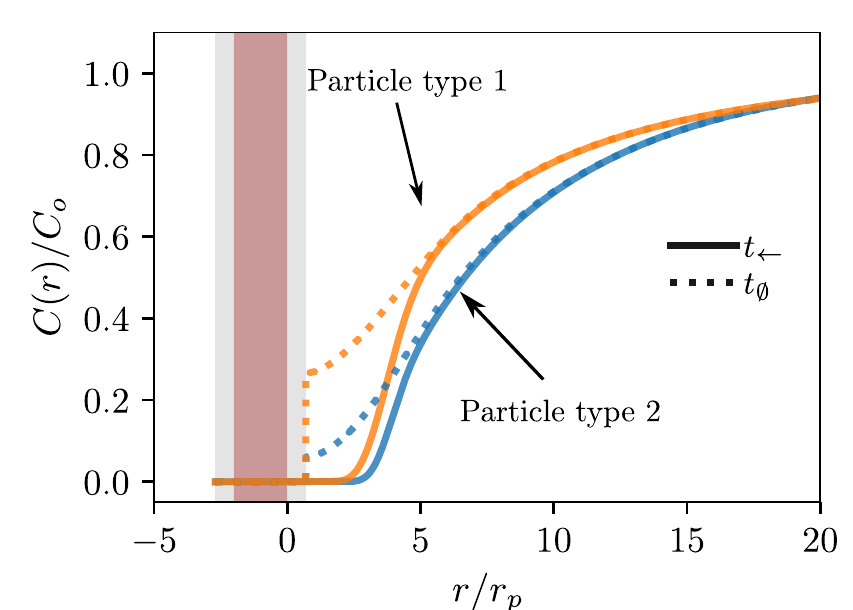}
\end{center}
\caption{Scaled concentration $C(r,t)/C_o$ \textit{vs} scaled radial distance $r/r_p$ for two particles with same mobility but two different diffusion coefficients (see text) under a time-varying field, after 100 cycles. Solid lines: at the end of the reverse phase of duration $t_{\shortleftarrow}=62.5\,\tau_p$. Dotted lines: at the end of the refill phase of duration $t_\emptyset= 196.9\,\tau_p$.The brown area marks the position of the wall/membrane with a thickness of $\ell_p$, and the gray bands are of thickness $R_p$.}
\label{Fig:sbc_rho_AC}
\end{figure}

Our KMC simulation scheme can easily be adapted to also include analyte field-driven drift across the nanochannel. The translocation process itself is simply 1D motion in the presence of a uniform field (of intensity $E_0$). The cross-sectional area of the channel is exactly the surface area of the absorber at $r=R_p$, and the field at that point is indeed $\approx E_0$. We thus extend the $x$-axis from $r=+R_p$ (where the absorber was located in the previous sections) to $r=-\ell_p-R_p$ (see Fig.~\ref{Fig:sbc_rho_AC}) and we don't use any entropic force when $r<R_p$. During the forward pulse of duration $t_\shortrightarrow$, we place the absorber at $r=-\ell_p-R_p$ on the \textit{trans}-side (this is replaced by a RBC during the other two phases in order to stop any leakage). During the recovery phase of duration $t_\emptyset$, a RBC is placed at $r=R_p$ so that analytes do not diffuse inside the nanochannel. These conditions strongly limit the contribution of non-field driven translocation of analytes. Finally, a source boundary is applied maintained at $R_b=60\,r_p$ throughout, and the system starts with a uniform concentration $C_o$.

Since $\lambda_e \sim Q$ and $\mu \sim DQ$, we must have $\lambda_e \sim 1/D$ in order to keep the mobility molecular size independent. In our simulations, we use $D_1 \! = \! 2D_2 \! = \! 2D$ and a field intensity $\lambda_e^{(2)}=50\,r_p=2\lambda_e^{(1)} \equiv \lambda_e$. In order to define the pulse durations in an unambiguous way, we need a time scale that does not depend on molecular size such as the \textit{cleanup time} $\tau_p$ needed to empty a zone of radius $R_p$,
\begin{equation}
\label{eq:taup}
    \tau_p = \tau_E(R_p,0)={R_p^3}/{3\lambda_e D},
\end{equation}
which is size-independent here since $\lambda_eD \sim \mu$. 

The choice of pulse parameters is guided by the three constraints presented above. We first test the following parameters: backward pulse $t_{\shortleftarrow}=62.5 \,\tau_p \ll \tau_\lambda$; refill time $t_\emptyset\approx r_{\shortleftarrow}^2/6D_1\approx 196.9\,\tau_p$; and capture time $t_\shortrightarrow=29.0\,\tau_p$. Figure~\ref{Fig:sbc_rho_AC} shows the steady-state concentration pattern for the two particles at the end of the reverse (solid lines) and refill (dotted lines) phases. Clearly, the depletion zone at the end of the $t_{\shortleftarrow}$ pulse is larger for particle (2): this is expected since the pulses are designed to keep them far from the nanopore. The molecules then diffuse toward the pore during the $t_\emptyset$ phase: obviously, the gap between the dotted lines near the pore must result in two different capture rates. In this particular case, the ratio $\overline{\rho}_1/{\overline{\rho}_2}$ between the mean capture rates of the two molecular species is $\approx2.5$. 

\begin{figure}[ht]
\begin{center}
\includegraphics[width=\columnwidth]{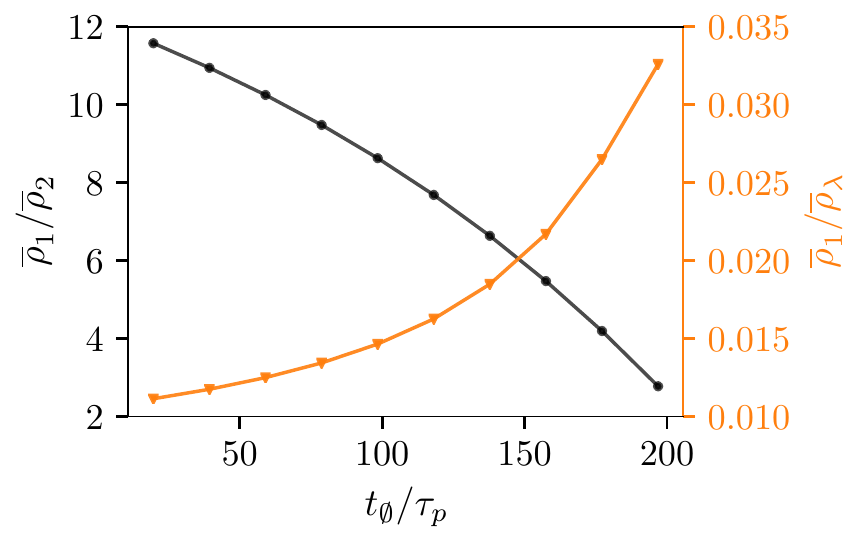}
\end{center}
\caption{The ratio $\overline{\rho}_1/{\overline{\rho}_2}$ of the mean capture rates of the two molecular species (\textcolor{black}{\large$\bullet$}), and the scaled mean capture rate $\overline{\rho}_1/\overline{\rho}_\lambda$ of the favored species (\textcolor{orange}{{$\blacktriangledown$}}), \textit{vs} the (scaled) recovery time $t_\emptyset/\tau_p$ for the system described in Fig.~\ref{Fig:sbc_rho_AC}.}
\label{Fig:sbc_rate_AC}
\end{figure}

The two capture rates and their ratio can both be modified by changing the refill time, as shown in Fig.~\ref{Fig:sbc_rate_AC}. The ratio of the mean capture rates increases by a factor of $\approx 6$ here; however, the mean capture rates also decrease ($\approx 3$ fold). An ideal device would have both a high capture ratio and a large capture rate, but in practice, this is not achievable: as usual in separation science, one has to choose between purity and speed. Optimizing the value of the other two time parameters for a given pair of same-mobility analytes is beyond the scope of this paper.

\section{Discussion and conclusion}
\label{sec:discussionandcon}

We have proposed a KMC algorithm that can efficiently map a spherically symmetric d-dimensional drift-diffusion problem onto a 1D biased random walk, and we successfully tested it using the standard problem of the diffusion-limited absorption of point-like particles by a sphere in both finite and infinite systems. The 1D projection greatly reduces the amount of memory needed to simulate large $d>1$ systems.

Our new KMC algorithm allows us to investigate both the short-time and steady-state dynamics of capture when a point-like external field is added. Our simulation results are in excellent agreement with the theoretical steady-state infinite-system concentration profiles and capture rates in 3D, further supporting the validity of the algorithm. Moreover, the short-time data agree with our theoretical estimates of the steady-state time $\tau_d \sim 1/D\lambda_e^2$ and of the size $r_d$ of the depletion region. We note that the depletion region is barely larger than the pore size, and that $\tau_d$ is too short to be observed in the 
lab.  

Interestingly, when the short-distance field is modified to take into account the finite width of the pore, the flatness of the (EAF) field near the pore creates a local traffic jam resulting in a substantially lower initial capture rate and a much increased steady-state time $\tau_d$. Furthermore, the plateau concentration is moved to larger distances $r_d$. Nevertheless, the same steady-state capture rate is eventually reached, and both $\tau_d$ and $r_d$ remain too small to be observable in typical experiments. Crucially, these results strongly suggest that one should not expect any useful depletion region or concentration patterns near the nanopore. 

When the system is a finite-size cavity, short time capture dynamics is similar to that observed in an infinite system, as one would expect. The region of the cavity that is beyond the capture radius $\lambda_e$ acts as a reservoir. As this reservoir is being slowly depleted, an outer depletion zone propagates inward from the cavity walls. The capture rate starts to decay from its initial steady state value when the front of the outer depletion region reaches the capture region at $r=\lambda_e$; however, since the time needed to reach this point increases rapidly with the cavity size ($\sim R_b^3$ if $R_b\gg \lambda_e$, which is normally the case), the capture rate can only be affected in small systems.

Of course, our KMC algorithm can also be used to simulate single particles dynamics. We thus revisited the reverse-polarity single-particle escape process that we introduced in our previous paper\cite{qiaoVoltagedrivenTranslocationDefining2019}. We again observe that the dynamics change from field-driven to diffusion-controlled at a distance $\lambda_e$, as expected from theory. Exploiting the fact that our 3D to 1D projection adds an entropic force to the equation of motion, we have proposed an improved approximation for the time-dependent mean trajectory $r(t)$ of the particles, in excellent agreement with the simulation data.

In the last part of the paper, we proposed a novel pulsed-field nanopore-based scheme to separate two different types of molecules with the same mobility but different diffusion coefficients. Section~\ref{section:separation} is intended to be a simple proof of concept showing that exploiting diffusivity differences might be possible. Clearly, one can change both the parameters, and even the shape, of the field pulses proposed here. For example, decreasing the recovery time can increase the ratio between the two capture rates, but at the cost of also decreasing the magnitude of both capture rates. In principle, one can also use an array of nanopores to enhance quantities, or a sequence of nanopores to enhance purity. 

\section*{Conflicts of interest}
There are no conflicts to declare.

\section*{Acknowledgements}
GWS acknowledges the support of both the University of Ottawa and the Natural Sciences and Engineering Research Council of Canada (NSERC), funding reference number RGPIN/046434-2013. LQ is supported by the Chinese Scholarship Council and the University of Ottawa.

\appendix
\section{The solution for a RBC and pure diffusion}
\label{sec:rbc_test}

With an ABC at $r=R_p$ and a RBC at $r=R_b$, the solution must satisfy the conditions $C(R_p,t)=0$ and $\partial_r C(r,t)|_{R_b}=0$.
We first define $u(r,t) = r C(r,t)$ and substitute this into eq. (\ref{eq:diff}) to obtain the simple differential equation
\begin{equation}
\label{Aeq:diffmax}
\frac{\partial u(r,t)}{\partial t}  = D \frac{\partial^2 u(r,t)}{\partial r^2}
\end{equation}
Using the separation of variables method with $u(r,t) = \phi(r) g(t) $, one obtains
\begin{equation}
\label{Aeq:diffmax_g}
g(t) = \exp(-tDk^2)
\end{equation}
and
\begin{equation}
\label{Aeq:diffmax_phi}
\phi(r) = A \cos(k (r-R_p)) + B \sin(k (r-R_p)),
\end{equation}
where $A$ and $B$ are constants and $k>0$.
The ABC at $r=R_p$ imposes that $A=0$ and the RBC at $r=R_b$ leads to eq.~\ref{eq:diffmax_Bn} for $k$. Note that eq.~\ref{eq:diffmax_Bn} has an infinite number of roots $k_n$ and must be solved numerically. When $n\gg1$, however, we find $k_n \rightarrow n \pi/ (R_b-r_p)$. Since the eigenfunctions $\phi_n$ are orthogonal, using the initial condition $u(r,t=0)=rC_o$, one can find the weights given eq.~\ref{eq:diffmax_Bn}.

\bibliography{Ref}
\end{document}